\documentclass{article}

\usepackage{arxiv}

\usepackage[utf8]{inputenc} 
\usepackage[T1]{fontenc}    
\usepackage{hyperref}       
\usepackage{url}            
\usepackage{booktabs}       
\usepackage{amsfonts}       
\usepackage{nicefrac}       
\usepackage{microtype}      
\usepackage{lipsum}		
\usepackage{graphicx}
\usepackage{doi}
\usepackage{subfigure}
\usepackage{siunitx}
\usepackage{todonotes}

\usepackage[bibstyle=numeric,
			citestyle=numeric-comp,
			natbib=true,
			backend=biber,
			maxcitenames=12,
		    mincitenames=10,
			sorting=none,]{biblatex}

\title{Exploring Perceived Vulnerability of Pedestrians: Insights from a Forced-Choice Experiment}

\author{ \href{https://orcid.org/0000-0001-7196-1245}{\includegraphics[scale=0.06]{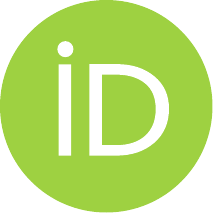}\hspace{1mm}Paul Geoerg}
    \\
	Vereinigung zur Förderung des Deutschen Brandschutzes e.V. \\
    Münster\\
    Germany\\
	\texttt{geoerg@vfdb.de} \\
	\And
 \href{https://orcid.org/0000-0003-3738-2869}{\includegraphics[scale=0.06]{im/orcid.pdf}\hspace{1mm}Ann Katrin Boomers}
    \\
	Forschungszentrum Jülich GmbH \\
    Jülich\\
    Germany\\
	\texttt{a.boomers@fz-juelich.de} \\
	\And
    Maxine Berthiaume
    \\
	University of Ottawa\\
    Ottawa\\
    Canada\\
	\texttt{mbert094@uottawa.ca} \\
 \And
 \href{https://orcid.org/0000-0001-7240-896X}{\includegraphics[scale=0.06]{im/orcid.pdf}\hspace{1mm}Maik Boltes}
    \\
	Forschungszentrum Jülich GmbH \\
    Jülich\\
    Germany\\
	\texttt{m.boltes@fz-juelich.de} \\
  \And
  \href{https://orcid.org/0000-0002-1669-3900}{\includegraphics[scale=0.06]{im/orcid.pdf}\hspace{1mm}Max Kinateder}
    \\
	National Research Council Canada  \\
    Ottawa\\
    Canada\\
	\texttt{max.kinateder@nrc-cnrc.gc.ca} \\
    }

\date{}

\renewcommand{\shorttitle}{}

\hypersetup{
pdftitle={Exploring Perceived Vulnerability of Pedestrians: Insights from a Forced-Choice Experiment},
pdfsubject={q-bio.NC, q-bio.QM},
pdfauthor={P. Geoerg, A. K. Boomers, M. Berthiaume, M. Boltes, M. Kinateder},
pdfkeywords={{Pedestrian dynamics, Heterogeneous crowd, Wheelchair user, Luggage, Vulnerability, Disability},
}
}

\begin{document}
\maketitle

\begin{abstract}
    Individual differences in mobility (e.g., due to wheelchair use) during crowd movement are not well understood. Perceived vulnerability of neighbors in a crowd could affect, for example, how much space is given to them by others. To explore how pedestrians perceive people moving in front of them, in particular, how vulnerable they believe them to be, we asked \SI{51}{} participants to complete a Two-Alternatives-Forced Choice task (2AFC) in an internet browser. Participants were shown pairs of images each showing a person and then asked to select the person who appeared more vulnerable to them. For example, participants would choose between a male person in a wheelchair and a female person carrying a suitcase. In total \SI{16}{} different stimuli (male vs female; no item/device, 1 suitcase, 2 suitcases, small backpack, large backpack, stroller, cane, and wheelchair), yielding $n(n-1)/2 = 120$ potential pairwise comparisons per participant. Results showed that wheelchair users appeared the most vulnerable and persons without any items/devices the least vulnerable. Persons carrying two suitcases were in the middle. These results informed the design of a main behavioral study (not reported here). 
\end{abstract}

\keywords{Pedestrian dynamics \and Accessibility \and Heterogeneous crowds \and Demographic change \and 2AFC \and Online study \and Perceived vulnerability \and Wheelchair user}

\section{Introduction}

This manuscript documents the background, rationale, procedure, and results of a pilot study designed to test the perceived vulnerability of pedestrians walking ahead of another person. The results of the pilot study informed the design of a larger behavioral study on pedestrian movement (not reported here, see \cite{Geoerg.2023f})

Individual differences in mobility (e.g., due to wheelchair use) during crowd movement are not well understood. This has been recognized in several \textit{review} publications, e.g.  \cite{Bukvic.2021, Carlsson.2022, Hostetter.2022}. While these have pointed out how, for example, certain functional limitations could affect pedestrian egress movement, they also report a lack of empirical data. Likely as a consequence, many engineering tools that aim to predict the evacuation performance of crowds are based exclusively on data from young adults without disabilities. However, previous work has found anecdotal evidence that pedestrians keep a larger distance from wheelchair users when walking in a crowd \cite{Geoerg.2022}. 

The perceived vulnerability of people walking next to each other (also called neighbors) in a crowd could potentially explain this effect on microscopic movement parameters (e.g., movement speed). More specifically, a relevant question that has not been answered is how the visibility of a disability (e.g., through recognition of an assistive device) or mobility-relevant properties (e.g., carrying heavy items, pushing strollers, etc.) shape how people see neighbors in a crowd. 

Individual pedestrians' behavioral reactions to visible mobility attributes may have cascading effects on micro- and macroscopic movement patterns in the crowd. Here, we explore two potential explanatory mechanisms: perceived vulnerability and perceived required space. That is, will people increase their interpersonal distance to wheelchair users because there is a social norm to be mindful of people who appear vulnerable or because wheelchair users simply appear to take up more space?

The purpose of this pilot study was to collect subjective impressions from participants via an online study that informed the main behavioral study (a group of participants moving through a bottleneck). Participants reported ratings of perceived vulnerability of pedestrians and wheelchair users with different mobility attributes using a Two-Alternatives-Forced-Choice (2AFC) paradigm \cite{Fechner.1860}. 

In a 2AFC paradigm, participants were shown two stimuli (e.g., images) and then need to select one. The 2AFC procedures allowed for generating unbiased responses and avoid certain response patterns compared to other procedures such as simple rating scales (e.g., anchoring) \cite{Fechner.1860, Bogacz.2006, Auerbach.1971}. 

\subsection{Research Questions}
\label{sec:research_questions}
The main research questions addressed in this pilot study were: 
\begin{enumerate}
    \item Are wheelchair users generally perceived to be more or less vulnerable than others?
    \item Are there differences/similarities compared to other mobility-related attributes?     
\end{enumerate}

\section{Methods}

\subsection{Design}
\label{sec:design}
In each trial, two images showing a person were presented. The participant's task was to select the person that appeared more vulnerable to them. More specifically, they were given the following instructions: "Imagine that you are walking behind each person while walking in a crowd. Please click on the person you would be more cautious around" (Figure \ref{fig:trial_screenshot}).\footnote{We queried feedback on optimal item formulation from an expert in ecological psychology and experimental design}  Participants then chose between images in which the following attributes were manipulated:
\begin{enumerate}
    \item Mobility attributes (8 levels): 
baseline (no mobility aid or luggage), 1 suitcase, 2 suitcases, small backpack, large backpack, stroller, cane, wheelchair
    \item Visible gender attributes of the person shown (male or female)
\end{enumerate}

This yielded a total of 16 unique stimuli (see \ref{sec:stimuli}). Stimuli were placed randomly either on the left or right side of the screen (see Figure \ref{fig:trial_screenshot}). Conditions in which two identical images were shown on the left and right side were excluded. In a 2AFC design, this translates into $n(n-1)/2 = 120$ potential pairwise comparisons. 

\subsection{Stimuli}
\label{sec:stimuli}
\begin{figure}
  \centering
  \subfigure[\label{fig:pilot_practice_stimuli}]{\includegraphics[width=0.6\linewidth]{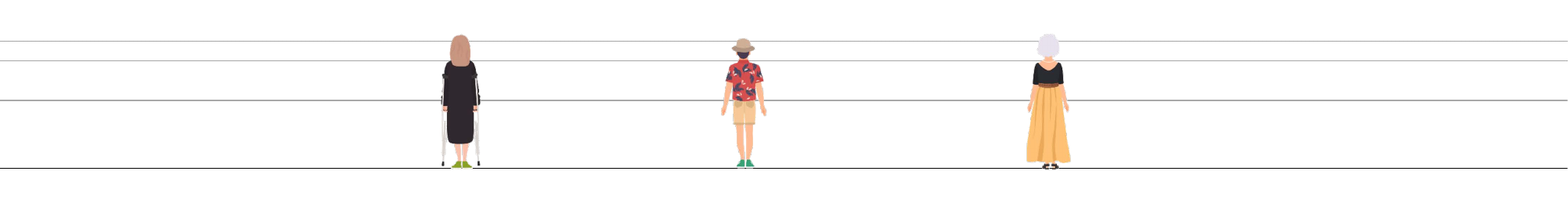}}\\
  \subfigure[\label{fig:stimuli_pilot}]{\includegraphics[width=0.75\linewidth]{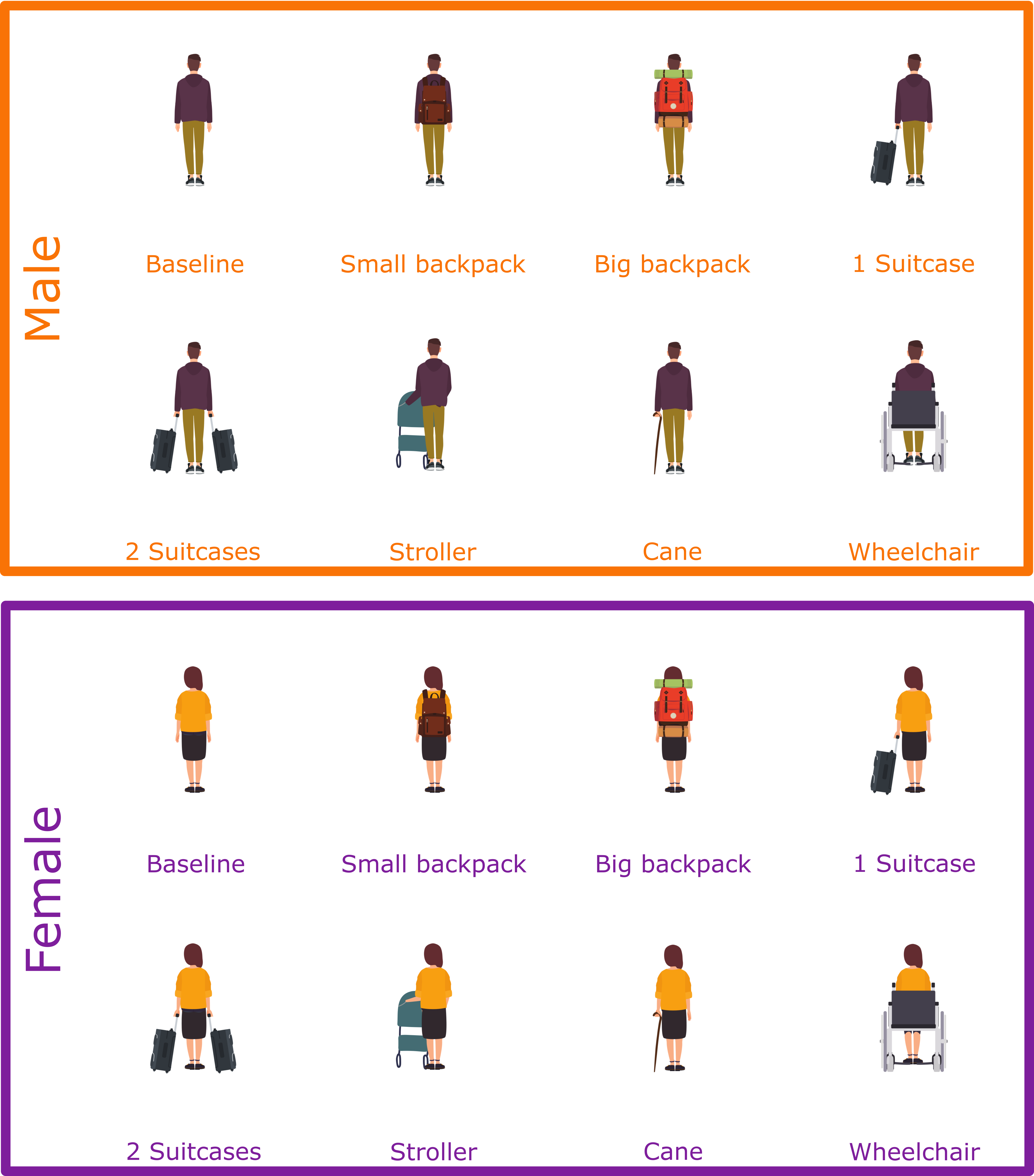}}\\
  \subfigure[\label{fig:trial_screenshot}]{\includegraphics[trim={0cm 0.8cm 0cm 1.5cm},clip, width=0.5\linewidth]{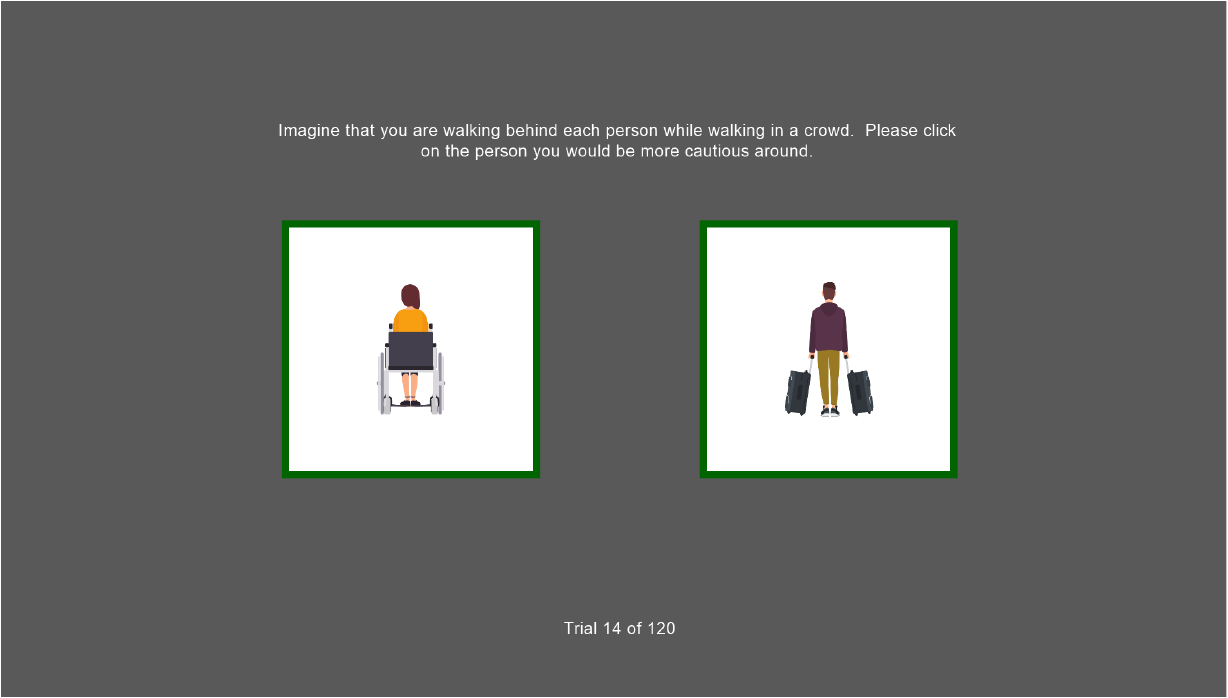}}
  \caption{Pilot study: (a) stimuli practice trials; (b) stimuli main trials; and (c) example screenshot of 2AFC task.}
\end{figure}

Sixteen stimuli were generated for the main trials of this study (Figure \ref{fig:stimuli_pilot}) as well as three stimuli for the practice trials (Figure \ref{fig:pilot_practice_stimuli}). Each stimulus was generated from a vector graphic based on two baseline models purchased from a commercial stock image database (shutterstock.com) and then loaded into an image editing software (Gimp v.2.10.32, www.gimp.org). The baseline images showed stylized views of an adult person, with either stereotypical male or female visual attributes. The "male" character had short hair, a jacket and pants. The "female" character had long hair and was wearing a skirt. Both characters were Caucasian. The baseline models were then modified for varying mobility profiles. Care was taken so that the basic appearance (e.g., posture, size) of the baseline characters did not change, except for the changes associated with the mobility profiles.
The three practice trial stimuli showed similar characteristics, however, their design differed so that no feature (e.g., assistive device) would match those of the main trials. They included a "female" character with crutches, a female character with grey hair, and an adult male character (Figure \ref{fig:pilot_practice_stimuli}).

\subsection{Procedure}
\label{sec:procedure}
\begin{figure}[ht]
    \center
     \includegraphics[width=1.0\linewidth]{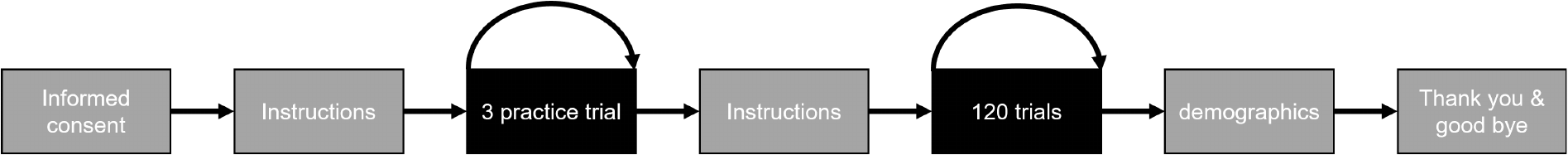}
     \\
     \caption{Pilot study structure and procedure.}
     \label{fig:procedure}
\end{figure}

The study was implemented using PsychoPy (v2021.1.4) and hosted online using Pavlovia.org (an online platform to run, share, and explore psychometric studies). Participants needed to use an internet-connected device (either computer, tablet, or phone) to access the study. The study procedure itself is structured in a sequential order (Fig\,\ref{fig:procedure}. Participants had to give informed consent before starting the study procedure, were instructed, practiced three exemplary trials to get familiar with the study design, and then started the 120 decision trials. Last, participants were asked for demographic information (age group, gender identity, prior experience with people with disabilities, and movement in crowds). The procedure was approved by the NRC Research Ethics Board (REB 2021-88). 

\subsection{Sample}
\label{sec:sample}

\begin{figure}
\centering
  \subfigure[\label{fig:age}]{\includegraphics[width=0.33\linewidth]{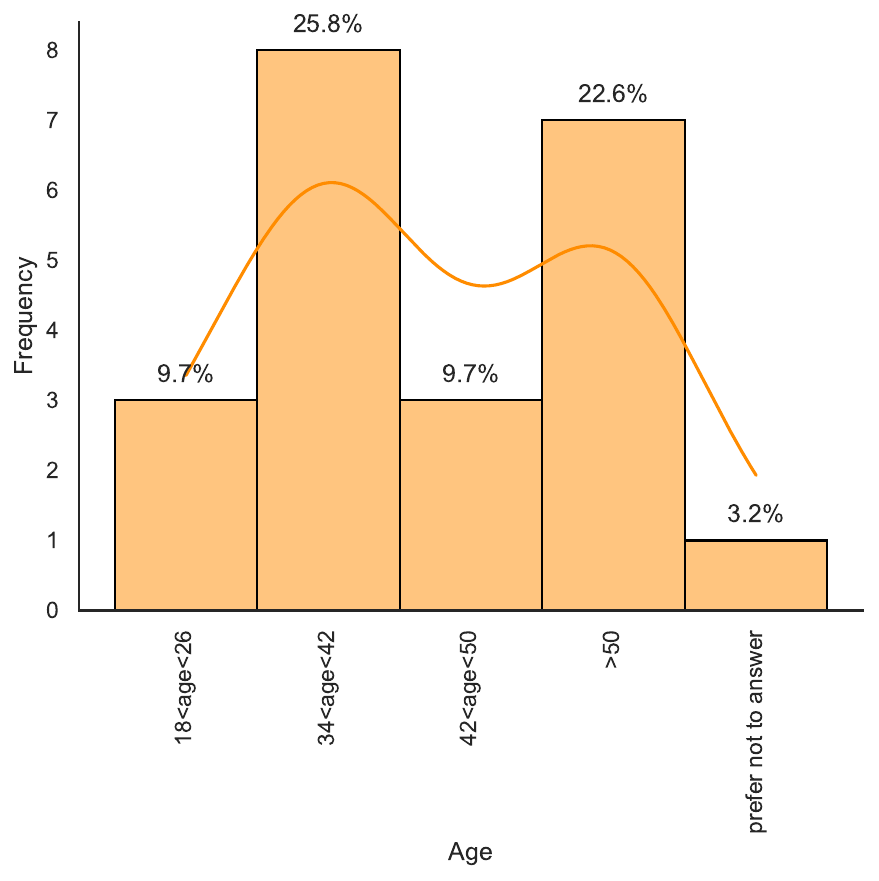}}
  \subfigure[\label{fig:gender}]{\includegraphics[width=0.33\linewidth]{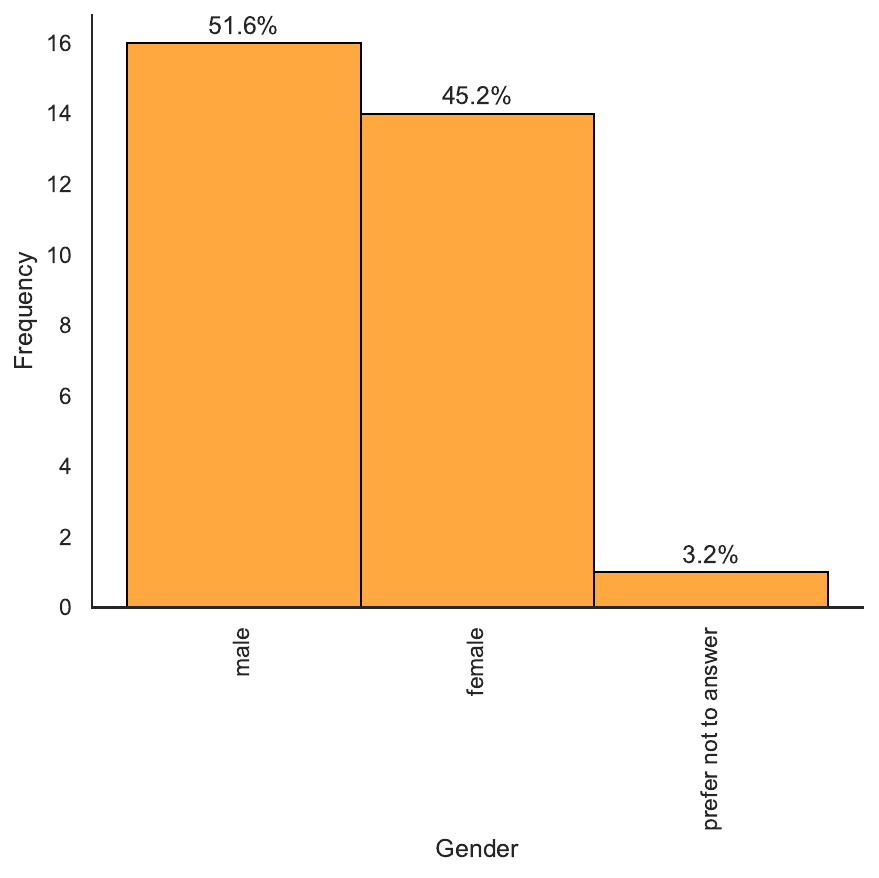}}
  \subfigure[\label{fig:disabilityexperienced}]{\includegraphics[width=0.33\linewidth]{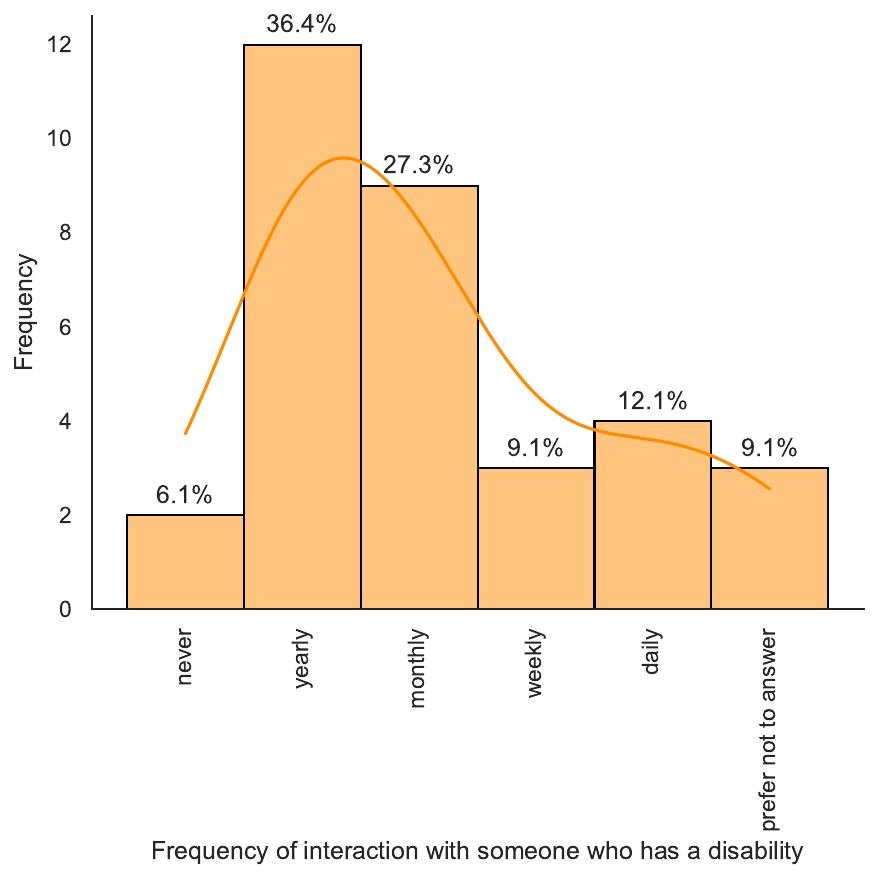}}
\caption{Distribution of (a) age, (b) gender, and (c) reported frequency of interaction with persons with disabilities. }
\label{fig:ageandgender}
\end{figure}

Participants were recruited via e-mail invitation and posting on social media. \SI{51}{} participants completed at least some of the main trials of the study. \SI{64.7}{\percent} ($n=\SI{33}{}$) of the participants provided a complete data set and provided demographic information. Two participants ($\approx \SI{6.1}{\percent}$) stated that they had a disability. Figure \ref{fig:ageandgender} shows the distribution of age, gender, and the amount of experience that participants had interacting with people with disabilities in their daily lives.


\clearpage
\newpage
\section{Results and Discussion}
\label{sec:results}

In this study, we investigated how visible mobility attributes, such as using a wheelchair, carrying luggage, or pushing a stroller affect the perceived vulnerability of pedestrians. 2AFC designs provide two metrics: Which image was selected (choice) and how long it took to make that decision. The former can provide direct answers to the research question (see section \ref{sec:research_questions}) and test for potential biases and stereotypical response patterns (e.g., a participant always clicking on the right stimulus). The latter can provide information on how difficult a decision was (the longer participants needed to decide between two stimuli, the harder the decision). In addition, response times can be used to filter non-credible responses (either too fast or too long, for a discussion in an adjacent area see \cite{Braw.2022}). 

\subsection{Data processing}
\label{sec:data_processing}
In order to exclude data with non-credible responses, we first filtered the data for response time by removing values above and below two standard deviations of the average data (response time $\bar{RT}=\SI{1.67}{\second} \pm \SI{7.62}{\second}$). Note that this approach has limitations (see \cite{Lachaud.2011} for a discussion). As a result, \SI{4.45}{\percent} (\SI{265}{}) from a total of \SI{5949}{} responses were removed from the original data set.

Next, we plotted histograms of the left and right clicks of participants to identify potential biases (see Figure \ref{fig:leftrightclicks}). None of the participants appeared to systematically prefer left to right and consequently, thus no  participants were excluded. 

\subsection{Choice}
\label{sec:results_choice}

\begin{figure}
\centering
  \includegraphics[width=0.75\linewidth]{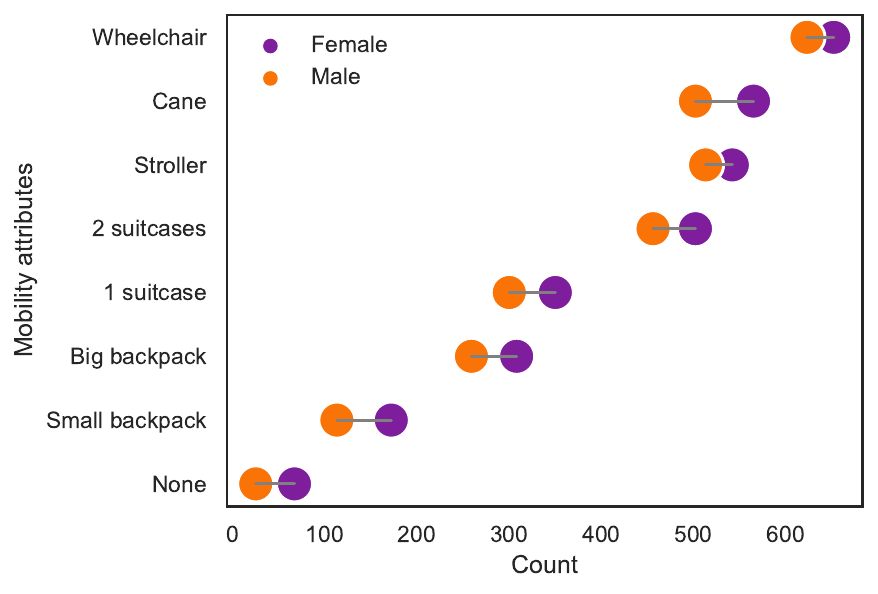}
  \caption{Absolute frequencies of each stimulus being selected.}
\label{fig:results_counts}
\end{figure}

Figure \ref{fig:results_counts} shows the absolute frequencies that each stimulus was selected. The following observations were made: Stimuli showing \ldots

\begin{enumerate}
    \item \ldots a person without any assistive device were consistently rated as the least vulnerable; 
    \item \ldots a person in a wheelchair appeared to be the most vulnerable, followed by those with a cane and stroller;
    \item \ldots persons with either one or two suitcases appeared to be in the middle;  
    \item \ldots persons carrying larger and smaller backpacks appeared less vulnerable than those carrying suitcases but more than a person without any assistive device or travel item;
    \item \ldots persons with stereotypically female attributes consistently appeared to be more vulnerable than those with male attributes. These differences were most pronounced for the stimuli showing \textit{cane} users, and smallest for stimuli showing \textit{wheelchair} and \textit{stroller} users. 
\end{enumerate}

This indicates that the perception of vulnerability scales with the space required by the item, i.e., a small backpack is perceived as less vulnerable than a large backpack, which is perceived as less vulnerable than a stroller. Similarly, the larger an item, the more vulnerable a person appears to be. Interestingly, this effect does not seem to be linear. For example, the differences in perception from small to big backpacks and from one to two suitcases were larger compared to the differences between other conditions. 

\subsection{Response times}
\label{sec:results_response_times}

\begin{figure}
\centering
  \subfigure[\label{fig:leftrightclicks}]{\includegraphics[width=0.45\linewidth]{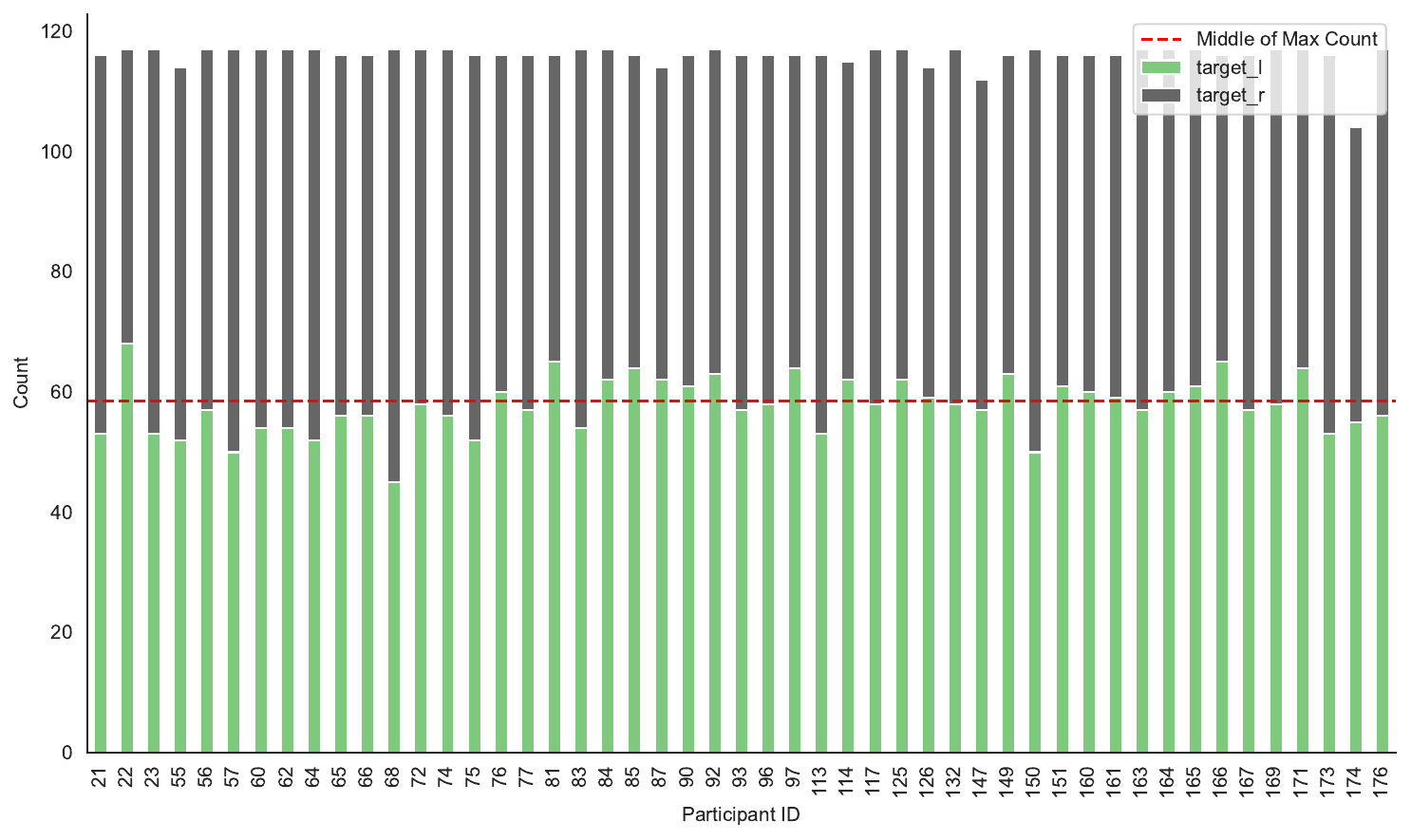}}
  \subfigure[\label{fig:response_time_by_stimulus}]{\includegraphics[width=0.45\linewidth]{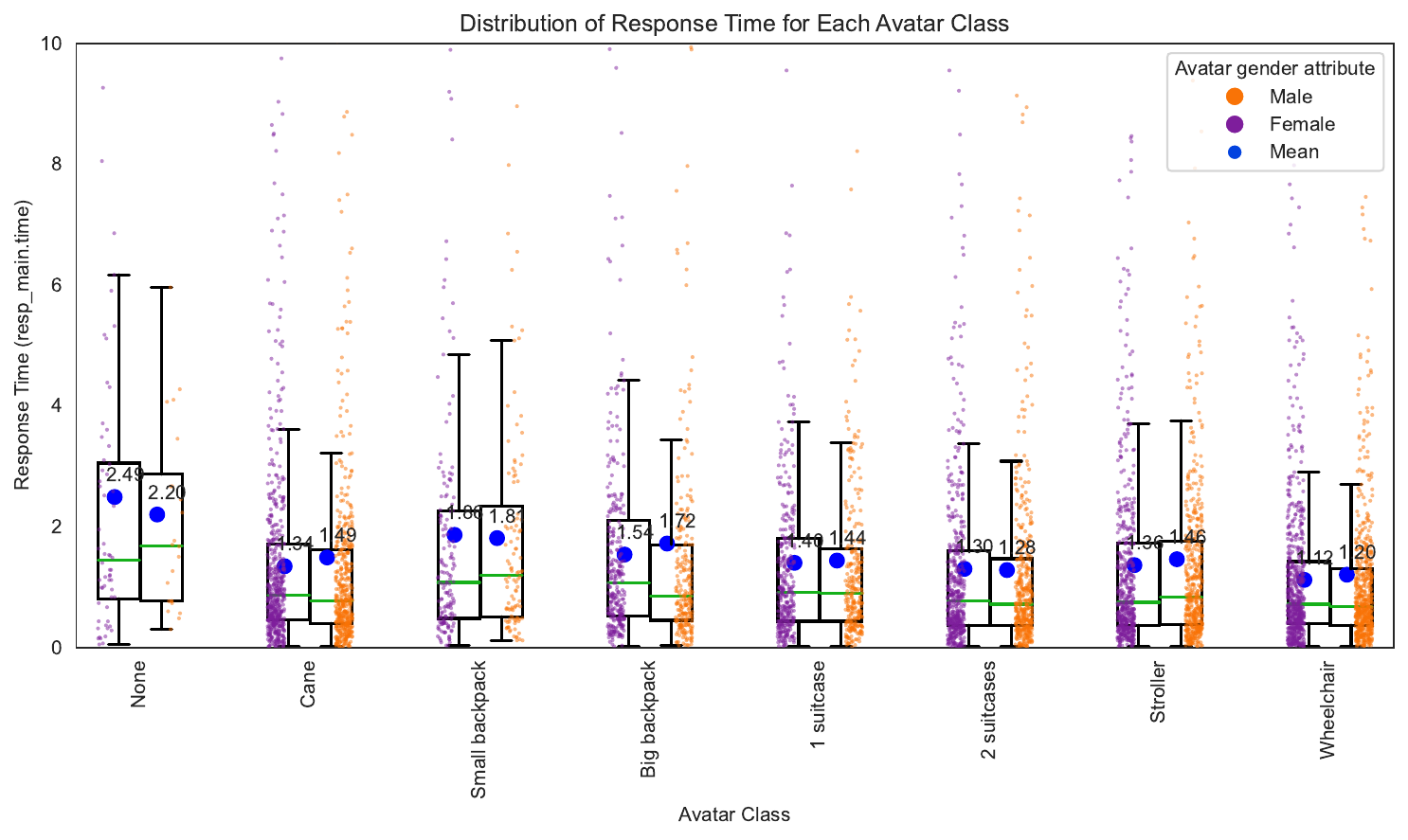}}
  \caption{(a) Histograms of left and right clicks for each participant ID, and (b) boxplots of response times for each stimulus, when selected.}
\label{fig:resptimes}
\end{figure}

Figure \ref{fig:response_time_by_stimulus} shows the response times for each stimulus, when selected. The average response time across all stimuli was $\SI{1.40}{} \pm \SI{1.79}{\second}$ (median $= \SI{0.81}{\second}$). Medians were consistently higher than averages. There were no noticeable differences in response times across stimuli (regardless of displayed gender attributes and items). Note that the number of data points varied notably because we only report the response times for the selected stimulus (see Figure \ref{fig:results_counts}). 

\begin{figure}
\centering
  \includegraphics[width=0.75\linewidth]{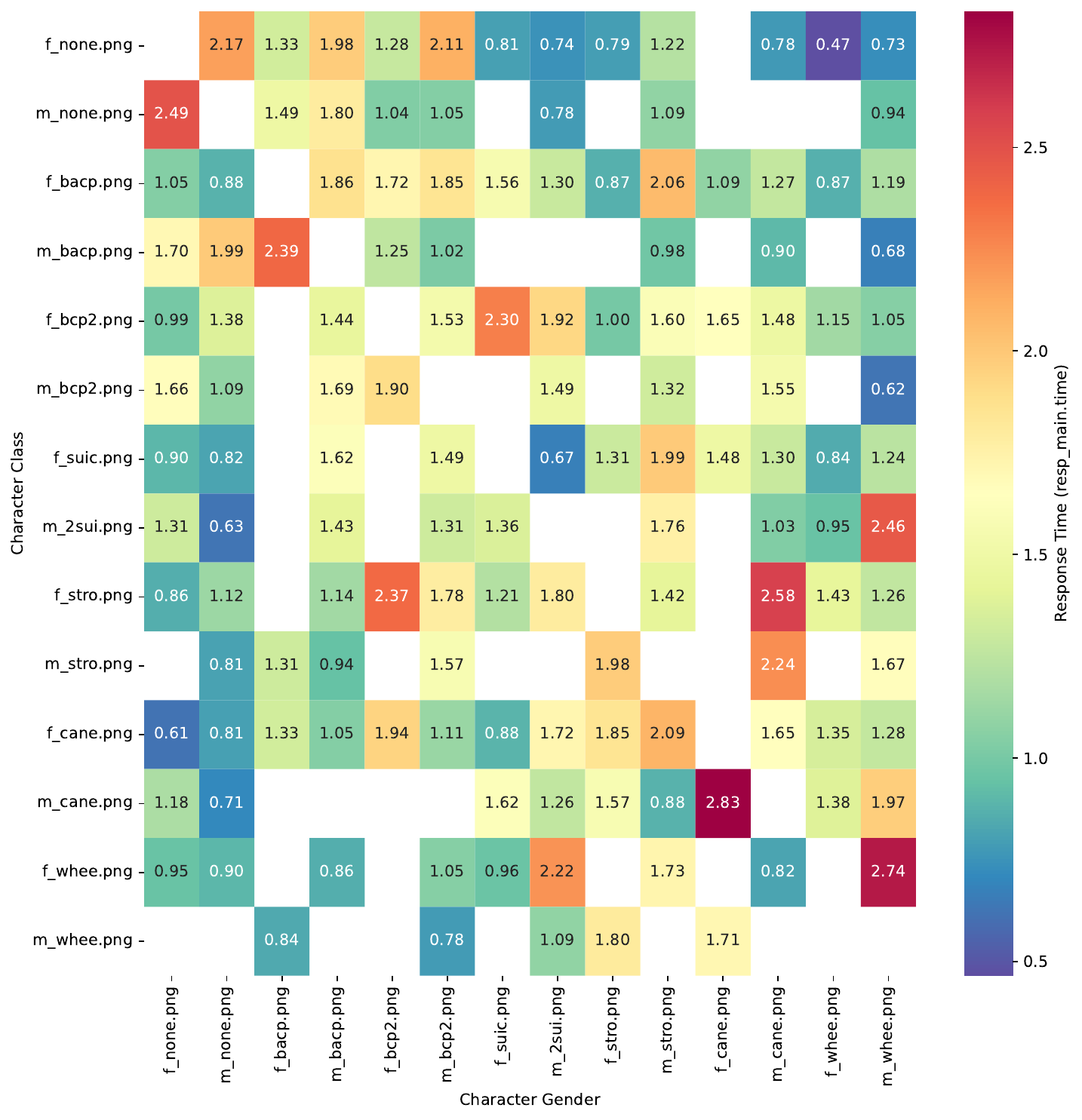}
  \caption{Heatmap of average response times as a function of pairwise comparisons. }
\label{fig:RT_LeftRightImage}
\end{figure}

Figure \ref{fig:RT_LeftRightImage} shows a heatmap of the average response time for each pairwise comparison. These data could indicate comparisons that were harder for participants to differentiate. The longest response times were reported for comparisons between male and female cane users (\SI{2.83}{s}), male with one suitcase and female with a large backpack (\SI{2.79}{s}), and male and female wheelchair users (\SI{2.74}{s}). The fastest response times were reported for comparisons between females with no items and female wheelchair users (\SI{0.47}{s}). This suggests the following patterns: 

\begin{itemize}
    \item Decisions between male and female figures were harder when they were otherwise similar in terms of mobility attributes.
    \item Decisions became easier when the contrast between items was larger (see, for example, the column for the male wheelchair user ($m_{whee}$) in Figure \ref{fig:RT_LeftRightImage}).
\end{itemize}

\section{Conclusions}
\label{sec:conclusion}
This study reports on a pilot experiment designed to provide insights and guidance for a larger behavioral study \cite{Geoerg.2023f}. The goal was to identify mobility profiles (i.e., attributes of people moving in a crowd) that differ in perceived vulnerability and space requirements. To this end, we asked participants to compare images showing characters that differed in their appearance, such as their stereotypical gender attributes (male vs female) or the kind of travel items (suitcases and backpacks) and assistive devices (cane and wheelchair) with them. 

We chose a 2AFC online study for this purpose and while this approach has clear limitations, a pattern could clearly be established, which answered the research questions and informed the design of the behavioral study: 
\begin{itemize}
    \item Pedestrians in wheelchairs consistently appeared to be more vulnerable than any other mobility profile (research question 1). The inverse was true for pedestrians without any other item/mobility device (research question 2).
    \item Other mobility-related attributes appeared to influence the perceived vulnerability as a function of their size, weight, and ease with which the items could be moved (e.g., characters with bag packs, in general, were seen as less vulnerable as those with suitcases and strollers). The only exception was the cane, which was likely interpreted as a clear sign of vulnerability, but to a lesser degree than the wheelchair (research question 2). 
    \item Within each category of items, characters that appeared to be female were consistently rated to be more vulnerable.
    \item The more similar the displayed stimuli were, the longer participants needed to decide. 
\end{itemize}

The present work has a number of limitations that should be considered. Firstly, we only reported on a selection of mobility profiles. The profiles were selected with the future behavioral study in mind, but the list is certainly not exhaustive. For instance, we did not vary the appearance of age (neither children nor seniors were displayed). Secondly, we did not measure perceived vulnerability directly, but asked participants to indicate the person they would be "more cautious" around. This approach was chosen given that the term 'vulnerable' may be interpreted differently based on the participants' own backgrounds. However, prior to testing, we solicited feedback on this formulation from native speakers of English who were experts in environmental psychology and experimental design. Finally, a large number of participants did not complete the full study; Consequently, the data set might be imbalanced given the amount of missing demographic information. In addition, the influence of the participant's own mobility profile (e.g., gender, age, or living with a disability) could not be investigated. 

However, we believe that the limitations did not prevent this work from achieving its goal of informing the design of a behavioral study. In the behavioral study, groups of participants will be asked to move together through a bottleneck; critically, the mobility attributes of two participants at the center of the group will be manipulated. In wheelchair conditions, two participants will be wheelchair users, in the luggage condition, two participants will be carrying two suitcases each, while in control conditions all participants will have similar mobility attributes.


\printbibliography

\end{document}